\documentclass[conference]{IEEEtran}
\IEEEoverridecommandlockouts

\usepackage{xurl}
\usepackage{cite}
\usepackage{amsmath,amssymb,amsfonts}
\usepackage{algorithmic}
\usepackage{graphicx}
\usepackage{textcomp}
\usepackage{xcolor}
\def\BibTeX{{\rm B\kern-.05em{\sc i\kern-.025em b}\kern-.08em
    T\kern-.1667em\lower.7ex\hbox{E}\kern-.125emX}}
\begin{document}

\title{Nishpaksh: TEC Standard-Compliant Framework for Fairness Auditing and Certification of AI Models
}

\author{\IEEEauthorblockN{Shashank Prakash}
\IEEEauthorblockA{
\textit{IIIT Delhi, New Delhi} \\
shashankp@iiitd.ac.in}
\and
\IEEEauthorblockN{Ranjitha Prasad}
\IEEEauthorblockA{
\textit{IIIT Delhi, New Delhi} \\
ranjitha@iiitd.ac.in}
\and
\IEEEauthorblockN{Avinash Agarwal}
\IEEEauthorblockA{
\textit{DOT, Govt. of India, New Delhi} \\
avinash.70@gov.in}
}

\maketitle

\begin{abstract}
The growing reliance on Artificial Intelligence (AI) models in high-stakes decision-making systems, particularly within emerging telecom and 6G applications, underscores the urgent need for transparent and standardized fairness assessment frameworks. While global toolkits such as IBM AI Fairness 360 and Microsoft Fairlearn have advanced bias detection, they often lack alignment with region-specific regulatory requirements and national priorities. To address this gap, we propose Nishpaksh, an indigenous fairness evaluation tool that operationalizes the Telecommunication Engineering Centre (TEC) Standard for the Evaluation and Rating of Artificial Intelligence Systems. Nishpaksh integrates survey-based risk quantification, contextual threshold determination, and quantitative fairness evaluation into a unified, web-based dashboard. The tool employs vectorized computation, reactive state management, and certification-ready reporting to enable reproducible, audit-grade assessments, thereby addressing a critical post-standardization implementation need. Experimental validation on the COMPAS dataset demonstrates Nishpaksh’s effectiveness in identifying attribute-specific bias and generating standardized fairness scores compliant with the TEC framework. The system bridges the gap between research-oriented fairness methodologies and regulatory AI governance in India, marking a significant step toward responsible and auditable AI deployment within critical infrastructure like telecommunications.
\end{abstract}


\section{Introduction}
The growing integration of Artificial Intelligence (AI) models into high-stakes decision-making systems across diverse domains, from hiring and healthcare to finance and law enforcement, has amplified concerns around bias and accountability. This urgency is particularly acute in the context of telecommunications, where AI is increasingly critical for network management, personalized service delivery, and resource allocation in emerging 6G paradigms. Biased AI within telecom can lead to unequal access to services, discriminatory pricing, or unfair resource distribution, directly impacting societal well-being and national digital inclusion goals. Consequently, there is an urgent need for transparent and systematic tools to evaluate and ensure fairness in these models. Several open-source frameworks have emerged globally to address this issue, such as IBM’s AI Fairness 360 \cite{bellamy2019ai}, Fairlearn by Microsoft \cite{weerts2023fairlearn} and What-If tool by Google \cite{article}. These tools offer a range of fairness metrics, bias mitigation algorithms, and visualization features that enable users to detect and analyze unfair behavior in AI systems.

India has a rapidly expanding AI ecosystem, which presents distinct challenges arising from its immense social, cultural, and demographic diversity. The data used to train AI models often reflects underlying social structures, where sensitive attributes such as caste, region, and economic background can unintentionally act as proxies for bias, making fairness evaluation particularly critical in the Indian context \cite{10885551}. Recognizing this, and aligning with national research priorities and the Bharat 6G vision, the Telecommunication Engineering Center (TEC) published the 'Standard for the Evaluation and Rating of Artificial Intelligence Systems' \cite{TEC57050}. This national standard establishes a structured three-step procedure that involves (a) classification of the risk of bias through a standardized questionnaire, (b) selection of the fairness metrics and thresholds, and (c) analytical bias testing to produce a composite Bias Index (BI) and Fairness Score (FS) via fairness metrics computation \cite{agarwal2023fairness}. Unlike general-purpose global frameworks, the TEC standard is explicitly designed to quantify fairness within India’s regulatory and demographic realities, making it a pivotal instrument for AI governance. However, while the TEC standard provides a well-defined framework for evaluating fairness, practical tools that operationalize this standard remain largely absent \cite{TEC57050}. 

Existing open-source libraries are designed for global contexts and do not align with the TEC’s procedures or India-specific data characteristics. Consequently, organizations seeking to comply with this national standard, especially within the telecom sector, often lack automated means to conduct bias assessments or generate standardized fairness reports. This gap highlights the pressing need for an indigenous, implementable auditing tool that translates the TEC guidelines into actionable, data-driven evaluations of AI fairness, thereby enhancing audit readiness and compliance for AI deployed in India \cite{10885551}.

In this work, we propose Nishpaksh, a fairness testing tool that implements the TEC standard by translating its guidelines into a web-based audit dashboard. The tool enables users to complete the three-step evaluation process and obtain fairness certification according to the national standard. This work is deeply rooted in the standardization lifecycle, both by addressing a critical post-standardization implementation gap of an existing national telecom standard and by proposing methodologies within the tool that could inform future pre-standardization efforts for AI risk grading and fairness assessment processes. To address potential gaps in user familiarity with the TEC standard, the tool includes a brief drop down explanation that outlines the key fairness requirements, ensuring that evaluators receive the necessary guidance before completing the audit process.

\noindent \textbf{Contributions:} Nishpaksh makes five key contributions towards operationalizing fairness assessment in India, aligning with national priorities for AI governance and standardization:
\begin{enumerate}
    \item It delivers the country’s first open-source tool that complies with the TEC standard, automating its three-step fairness evaluation methodology and thereby bridging a critical gap in the practical implementation of national AI governance standards within the telecom domain and beyond.
    \item It enhances transparency through structured logging, traceable metric selection and computation, and standardized report generation, directly supporting post-standardization evaluation and interoperability goals.
    \item It introduces contextual sensitivity by incorporating India-specific demographic attributes and proxy bias detection mechanisms into the evaluation workflow.
    \item It advances audit readiness by generating reports aligned with the TEC’s certification template, thereby enabling both self assessment and independent auditing of deployed AI models.
    \item It proposes standardization-ready processes for AI fairness assessment, particularly through its structured risk classification questionnaire, which can inform the development of broader frameworks for AI application risk grading in future telecom and other industrial standards.
\end{enumerate}

Collectively, these contributions bridge the gap between research oriented fairness frameworks and standardized AI governance practices in India. In the sequel, we describe the design and methodology adopted towards the proposed fairness assessment tool, and provide empirical results validating its performance and applicability.

\section{Related Works}

The field of algorithmic fairness has advanced through a range of bias detection and mitigation frameworks, each addressing fairness from distinct architectural standpoints. IBM AI Fairness 360 (AIF360) is among the most comprehensive, offering over seventy bias metrics and nine mitigation algorithms across pre, in, and post-processing stages \cite{bellamy2019ai}. Its modular architecture, introducing standardized Metric and Explainer classes, set early benchmarks for fairness evaluation. However, AIF360’s reliance on the IBM Cloud and global orientation limit its adaptability to region-specific governance standards, reducing its suitability for localized compliance and audit requirements.

Microsoft’s Fairlearn emphasizes computational evaluation \cite{weerts2023fairlearn}. Its MetricFrame API enables disaggregated performance analysis to identify allocation harms, and its strong interoperability with Python’s ML ecosystem supports industrial deployment. Yet, it lacks alignment with formal certification or regulatory frameworks.

Google’s What-If Tool (WIT) focuses on visualization-driven exploration and counterfactual analysis via TensorBoard and Jupyter integrations \cite{article}. While accessible to non-technical users, its exploratory emphasis limits reproducibility and audit readiness.

Several other toolkits address fairness from complementary perspectives: LiFT (LinkedIn Fairness Toolkit) enables large-scale fairness auditing in distributed machine learning systems \cite{vasudevan2020lift}, the Fairness R package provides statistical bias metrics for analytical workflows \cite{smith2025fairmetrics}, and Aequitas offers visualization and reporting tools for intuitive analysis of group disparities \cite{saleiro2018aequitas}.

\noindent \textbf{Novelty:}~These frameworks share fundamental gaps: (1) absence of regulatory alignment with national certification standards, (2) limited lifecycle coverage beyond model level evaluation , (3) generic threshold determination without context specific calibration, and (4) lack of certification ready reporting mechanisms. While existing tools are effective for algorithmic evaluation, they fall short of connecting technical assessments with country-specific regulatory compliance requirements. Table~\ref{tab:Comparison} presents a concise comparison of these tools alongside Nishpaksh.

\begin{table*}[htbp]
\label{tab:Comparison}
\centering
\caption{Comparative overview of major fairness assessment toolkits}
\begin{tabular}{|p{3cm}|p{3cm}|p{5.5cm}|p{5.5cm}|}
\hline
\textbf{Tool} & \textbf{Core Focus} & \textbf{Strengths} & \textbf{Limitations} \\ \hline

\textbf{IBM AI Fairness 360 (AIF360)} & 
Comprehensive fairness evaluation and mitigation &
Strong interpretability via Metric and Explainer classes and extensible integration with ML pipelines. &
Global orientation with limited region-specific regulatory alignment (e.g., TEC) and relatively high computational and setup overhead. \\ \hline

\textbf{Fairlearn} & 
Fairness performance trade offs and disaggregated evaluation &
MetricFrame API enables group-wise fairness metrics, integrates seamlessly with Python ML stack (scikit-learn, PyTorch, TensorFlow) and transparent fairness–accuracy trade-off visualization. &
Limited GUI, mainly supports classification tasks and lacks built-in mitigation beyond reweighting and constraints. \\ \hline

\textbf{What-If Tool (Google)} & 
Interactive and visual fairness exploration &
Enables no-code counterfactual and subgroup analysis within TensorBoard and Jupyter, promotes model interpretability and accessibility for non-programmers. &
Bound to TensorFlow and Google Cloud ecosystem, minimal extensibility for custom metrics, focuses more on interpretability than quantitative mitigation. \\ \hline

\textbf{Nishpaksh (Proposed)} &
Standardized, regulation-aligned fairness evaluation &
Unifies bias survey, preprocessing, metric computation, threshold calibration, and visualization in a single dashboard, model-agnostic and compliant with TEC standard. &

Currently optimized for tabular ML models, planned extensions for image and text based fairness evaluation. \\ \hline

\end{tabular}
\label{tab:fairness_tool_comparison}
\end{table*}

\section{TEC Standard and Evaluation}
In this section, we describe the lifecycle based assessment and risk quantification adopted by the TEC standard. Subsequently we describe the process of contextual threshold determination and inference and evaluation as proposed in the standard. 

\subsection{Lifecycle based assessment and risk quantification
}

The proposed \textit{Nishpaksh} tool employs a structured evaluation methodology encompassing seven key domains aligned with the TEC standard’s multi-dimensional view of bias.Each domain is rated on a five-point ordinal risk scale: very low (1), low (2), medium (3), high (4), and very high (5).

\textbf{Data Domain Assessment:}~ Evaluates data collection, preprocessing, annotation, and splitting stages. Key criteria include historical bias inheritance, representational adequacy, dataset completeness, and annotation consistency across protected groups.

\textbf{Model Domain Assessment:}~ Covers architecture design, training methodology, and performance monitoring. It examines pre-trained model validation, fairness implications of objectives, feature selection rationale, hyperparameter tuning, and group-wise performance consistency, distinguishing between newly introduced and inherited biases.

\textbf{Pipeline and Infrastructure Assessment:}~ Examines robustness against data leakage, uncertainty handling, and fairness implications of optimization choices balancing efficiency and equity. It identifies whether computational priorities override fairness considerations.

\textbf{Interface and Integration Assessment:}~ Reviews accessibility and user experience design to detect interface-level disparities or barriers affecting specific demographic groups.

\textbf{Deployment Analysis:}~ Assesses distribution shifts between training and production data, temporal stability of inference, and potential dataset drift causing uneven performance across groups.

\textbf{Human-in-the-Loop Assessment:}~ Evaluates human oversight in decision-making and outcome monitoring, identifying whether human intervention mitigates or amplifies demographic bias.

\textbf{System-Level Assessment:}~ Provides an end-to-end analysis through error rate disparity and user-journey bias mapping, capturing fairness implications beyond isolated components.

The framework operationalizes the TEC standard’s distinction between process and technical factors. Each assessment item is tagged accordingly, enabling weighted aggregation and targeted mitigation strategies.

The risk calibration criteria are defined as follows:
\begin{enumerate}
\item \texttt{High}: Models are highly vulnerable to adversarial manipulation and exhibit significant disparate error rates.
\item \texttt{Medium-High}: Basic assessment protocols exist with partially implemented oversight mechanisms.
\item \texttt{Medium}: Standard methodologies are applied without explicit bias awareness, focusing mainly on performance.
\item \texttt{Low}: Bias-aware methodologies are adopted with systematic fairness considerations during development.
\item \texttt{Very Low}: Bias is proactively mitigated at all stages with validated practices and well-calibrated systems.
\end{enumerate}

\subsection{Contextual Threshold Determination}
In this second stage of model evaluation, users specify the AI model type (e.g., machine learning, deep learning) and task category (e.g., regression, classification). It comprises three components:

\textbf{Model Accountability:}~ Standardized questions ensure clarity about the model’s purpose, design, and intended use.

\textbf{Sector Requirements:}~ Supports domain-specific selection of fairness metrics and thresholds reflecting context-dependent risks (e.g., healthcare vs. recruitment).

\textbf{Transparent Reporting:}~ Ensures stakeholder interpretability through clear metric reporting and stepwise documentation.

\subsection{Inference and Results Architecture}
This final evaluation stage includes:

\textbf{Quantitative Evaluation Layer:}~ Performs data harmonization (inputs with labels and sensitive attributes), computes performance metrics (accuracy, recall, specificity, precision, etc.), and evaluates fairness metrics (e.g., Statistical Parity Difference, Equal Opportunity Difference, Theil Index). Bootstrap resampling provides $95\%$ confidence intervals for robust inference.

\textbf{Visualization Layer:}~ Offers model summaries, uncertainty plots, group-wise disparity panels, and fairness–performance trade-offs.

\textbf{Results Section:}~ Computes standardized indices such as the Bias Index (BI) and Fairness Score (FS) for benchmarking and quantify bias arising from multiple sensitive attributes.

Outputs from all stages are compiled into a structured report including: (a) \textit{Summary} — model metadata and risk category, (b) \textit{Tabulation} — sub-scores and composite risk score, and (c) \textit{Detailed Analysis} — complete responses, visualizations, and findings.

This design ensures metric reproducibility, transparent documentation, and readiness for both developer self-certification and third-party audit validation.

\section{Design and Architecture of Nishpaksh}
The \textit{Nishpaksh} fairness evaluation tool is built on a modular architecture that integrates persistent session control, reactive computation, and concurrent data handling within the \textit{Streamlit} framework. Its core design incorporates JSON-based session serialization for seamless workflow resumption, decorator-level caching to eliminate redundant metric computations, and a parallelized data processing pipeline. Fairness metrics are derived using vectorized group-partitioned operations with bootstrap-based confidence estimation, while a reactive dependency framework ensures automatic propagation of updates across dependent components upon state changes. The system architecture is organized into five sequential stages: (a) survey intake and risk quantification, (b) metric selection and threshold specification, (c) proxy-aware feature engineering, (d) model inference with uncertainty estimation, and (e) threshold-driven composite scoring.

\begin{figure}[htbp]
    \centering
    \includegraphics[width=0.85\linewidth]{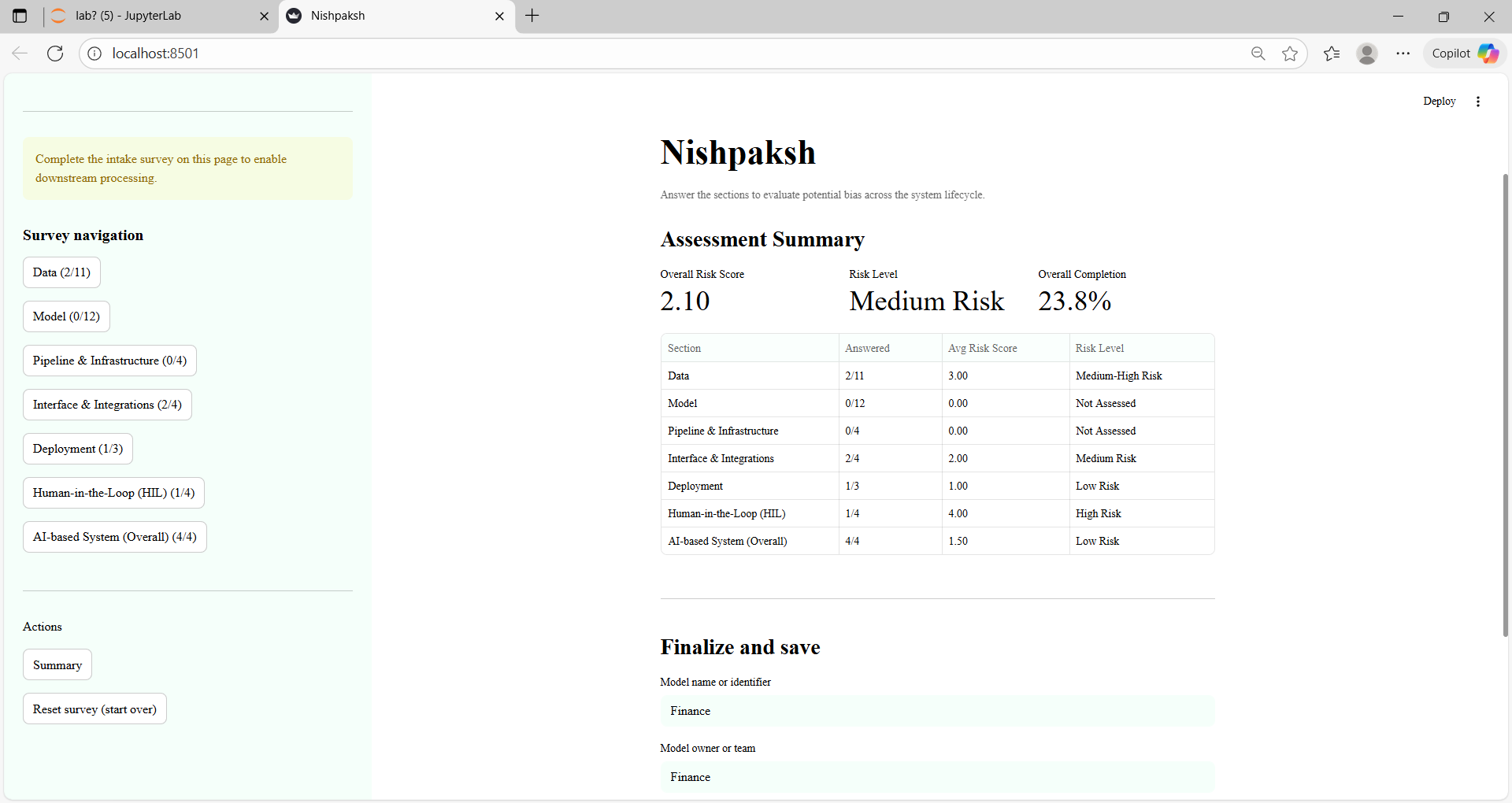}
    \caption{User interface of the \textit{Nishpaksh}  showing the survey intake and configuration panels.}
    \label{fig:nishpaksh_ui}
\end{figure}

\begin{figure}[htbp]
    \centering
    \includegraphics[width=0.90\linewidth]{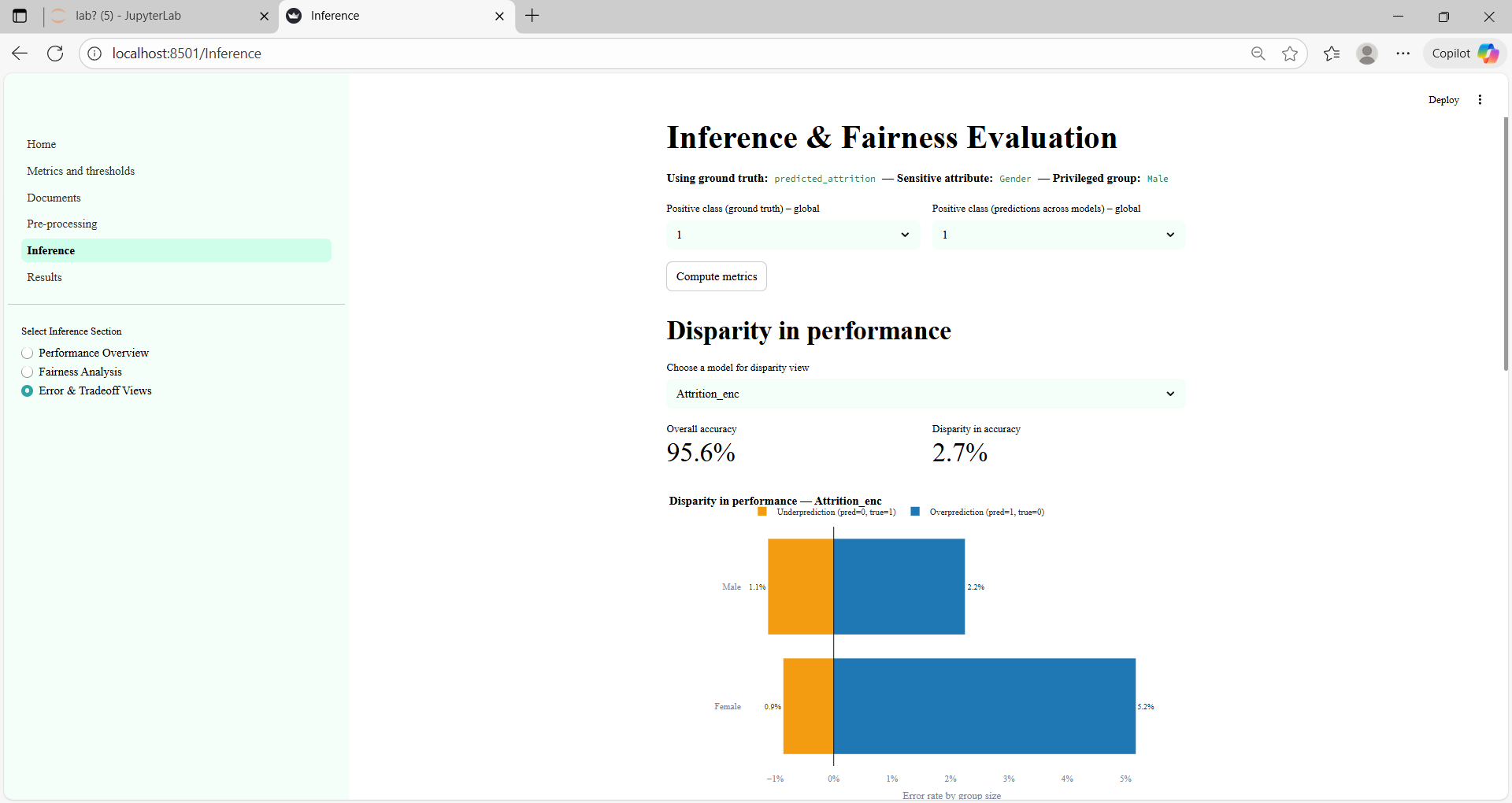}
        \caption{Inference and Fairness Evaluation interface of the \textit{Nishpaksh} showing model wise disparity analysis for sensitive attribute.}
    \label{fig:inference_section}
\end{figure}

The computational backend integrates high-cost tasks such as cross-validation, metric computation, and resampling, supported by multi-threaded execution to reduce memory overhead. At each stage, intermediate results are appended to source data frames, enabling efficient state tracking. Workflow continuity is maintained through JSON checkpoints, allowing seamless interruption and resumption; this also facilitates report generation at the end of the audit. In Fig.~\ref{fig:nishpaksh_ui} and Fig.~\ref{fig:inference_section}, we provide representative screenshots of the survey, configuration panels and inference interface of the tool.

\section{Experiments and Results}

\subsection{Dataset Description}
The proposed fairness assessment framework was evaluated using the COMPAS recidivism dataset, a benchmark dataset in algorithmic bias research. It contains demographic, behavioral, and criminal history information with a binary outcome variable \texttt{two\_year\_recid}. The sensitive and the nonsensitive features are as follows:
\begin{itemize}
    \item Non sensitive: age, priors\_count
    \item Sensitive: race (Caucasian=1, Non-Caucasian=0), sex (Male=1, Female=0)
    \item Target: two\_year\_recid (0=no recidivism, 1=recidivism)
\end{itemize}

\begin{figure}[htbp]
    \centering
    \includegraphics[width=0.45\textwidth]{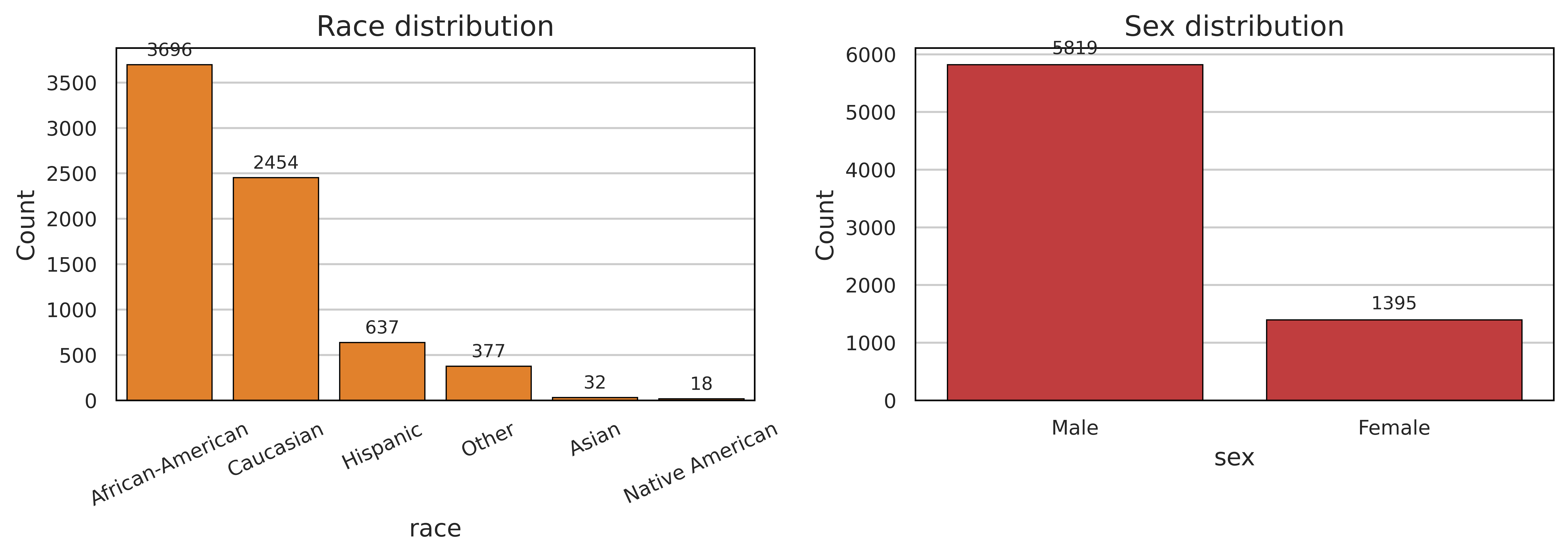}
    \caption{Distribution of sensitive attributes in the COMPAS dataset.}
    \label{fig:data_dist}
    \vspace{-3mm}
\end{figure}

In our experiments, these attributes are used to detect both data imbalance and downstream predictive bias. In Fig.~\ref{fig:data_dist}, we illustrate the distribution of sensitive attributes. We observe that the dataset exhibits significant racial imbalance, 
with a higher representation of African-American defendants as compared to other racial groups. This imbalance provides an empirical basis 
for observing biased outcomes in the trained models.

\subsection{Experimental Setup}
Three logistic regression models were designed to represent varying levels of bias exposure:
\begin{itemize}
\item \textbf{Baseline (Fair):}~ Trained with fairness constraints to enforce demographic parity.
\item \textbf{Race Bias Model:}~ Trained using race-sensitive features without applying fairness constraints.
\item \textbf{Gender Bias Model:}~ Trained with gender-related attributes and no constraint enforcement.
\end{itemize}
Each setup captures a distinct fairness exposure scenario. Metrics were evaluated on unseen validation data, and the developed fairness dashboard automatically generated visual reports to ensure transparency and reproducibility.

\subsection{Fairness Metrics and Formulations}
To quantitatively assess bias, six widely recognized group fairness metrics were employed: Disparate Impact (DI) \cite{10.1145/2783258.2783311}, Statistical Parity Difference (SPD), Normalized Disparate Impact (NDI), Equal Opportunity Difference (EOD), Average Odds Difference (AOD), and Equalized Odds (EO) \cite{10.1145/3219819.3220046}. These measures are computed from the joint distribution of predicted outcomes $(\hat{Y})$, ground truth labels $(Y)$, and the sensitive attribute $(A)$, where $A=1$ and $A=0$ correspond to the privileged and unprivileged groups, respectively.
The formal definitions are as follows:

\noindent\textbf{Statistical Parity Difference (SPD):}~Quantifies the absolute difference in the probability of favorable outcomes between groups. 
\begin{equation}
SPD = P(\hat{Y}=1|A=1) - P(\hat{Y}=1|A=0)
\end{equation}
A perfectly fair model yields $SPD = 0$.

\noindent\textbf{Disparate Impact (DI):}~Evaluates the relative likelihood of favorable outcomes; values close to 1 indicate demographic parity.
\begin{equation}
DI = \frac{P(\hat{Y}=1|A=1)}{P(\hat{Y}=1|A=0)}
\end{equation}

\noindent\textbf{Normalized Disparate Impact (NDI):}~A normalized variant of DI that rescales its range to $[-1,1]$, where $NDI=0$ denotes parity, given as $NDI = DI - 1$.

\noindent\textbf{Equal Opportunity Difference (EOD):}~Captures disparity in true positive rates across groups, with $EOD=0$ representing equal opportunity.
\begin{equation}
EOD = P(\hat{Y}=1|Y=1,A=1) - P(\hat{Y}=1|Y=1,A=0)
\end{equation}

\noindent\textbf{Average Odds Difference (AOD):}~Represents the mean deviation in false positive and true positive rates between groups. Smaller magnitudes imply higher fairness and is given as
\begin{equation}
AOD = \frac{1}{2}\left[(FPR_{A=1} - FPR_{A=0}) + (TPR_{A=1} - TPR_{A=0})\right]
\end{equation}

\noindent\textbf{Equalized Odds (EO):}~Assesses joint equality in opportunity and error rates. Perfect fairness corresponds to $EO=0$, and is given as
\begin{equation}
EO = |FPR_{A=1} - FPR_{A=0}| + |TPR_{A=1} - TPR_{A=0}|
\end{equation}

To summarize fairness across multiple metrics, a composite scoring mechanism was designed using the Bias Index (BI) and Fairness Score (FS). 
The Bias Index is defined as~\cite{agarwal2023fairness}:
\begin{equation}
BI_i = \sqrt{\frac{1}{n} \sum_j (M_{ij} - M'_j)^2}
\end{equation}
where $M_{ij}$ represents the $j$-th fairness metric for the $i$-th model, and $M'_j$ corresponds to the same metric for the baseline fair model. 

The Fairness Score (FS) is defined as~\cite{agarwal2023fairness}: 
\begin{equation}
FS = 1 - \sqrt{\frac{1}{m} \sum_{i=1}^{m} (BI_i)^2}
\end{equation}
where $m$ denotes the number of sensitive attributes considered. 

\subsection{Fairness Metrics Results and Visualizations}
\begin{figure}[htbp]
    \centering
    \includegraphics[width=0.48\textwidth]{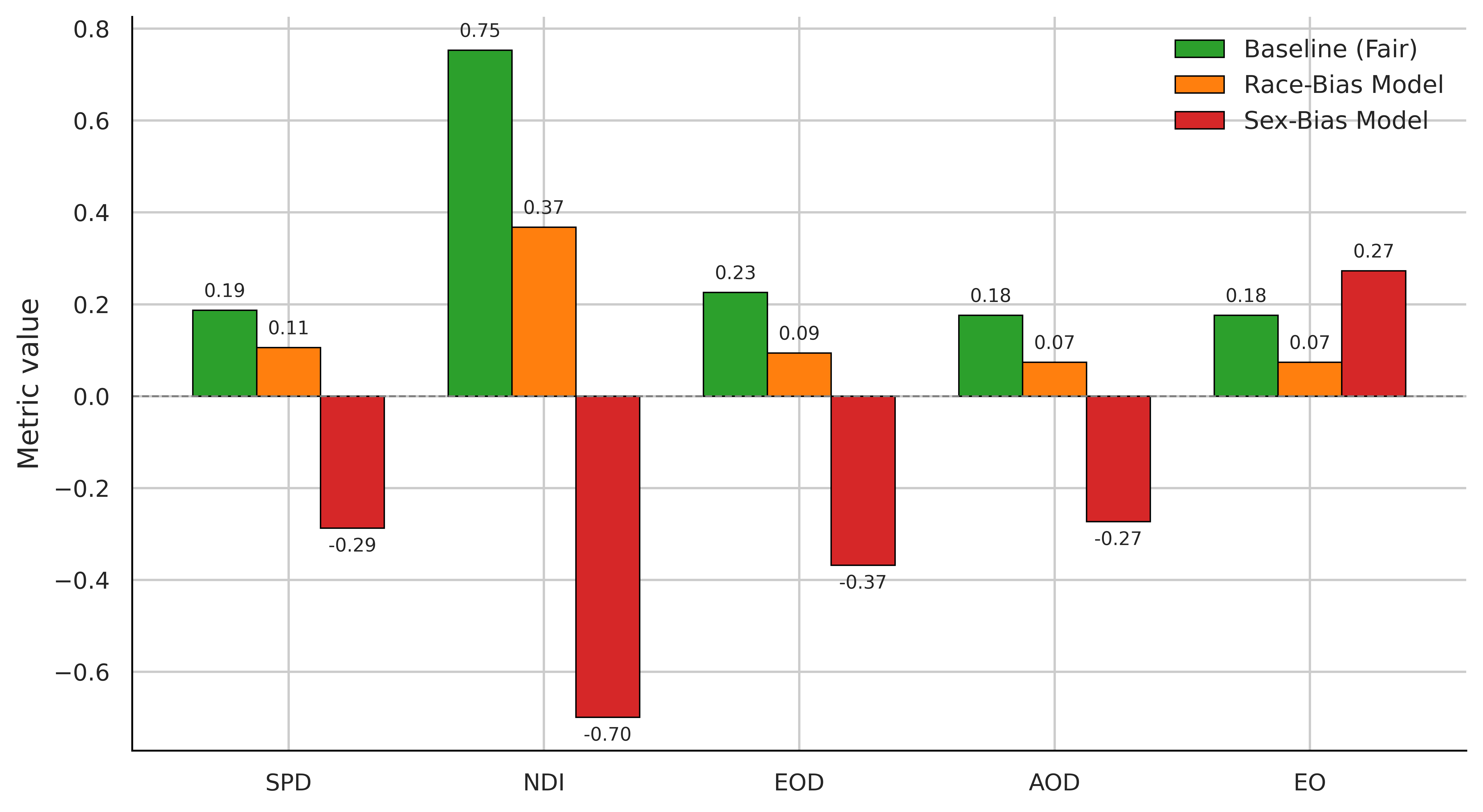}
    \caption{Fairness metric comparison across bias exposure conditions.}
    \label{fig:metrics}
\end{figure}

Table~\ref{tab:fairness_metrics} presents the computed fairness metrics for all model configurations, while Figure~\ref{fig:metrics} visualizes these variations for direct comparison. Lower magnitudes of SPD, EOD, and AOD correspond to reduced bias levels. The baseline (fair) model demonstrates close to ideal case parity across groups, whereas the race and gender biased models exhibit noticeable degradation, particularly in NDI and EOD, confirming attribute specific bias propagation within the learned representations.

\begin{figure}[htbp]
    \centering
    \includegraphics[width=0.45\textwidth]{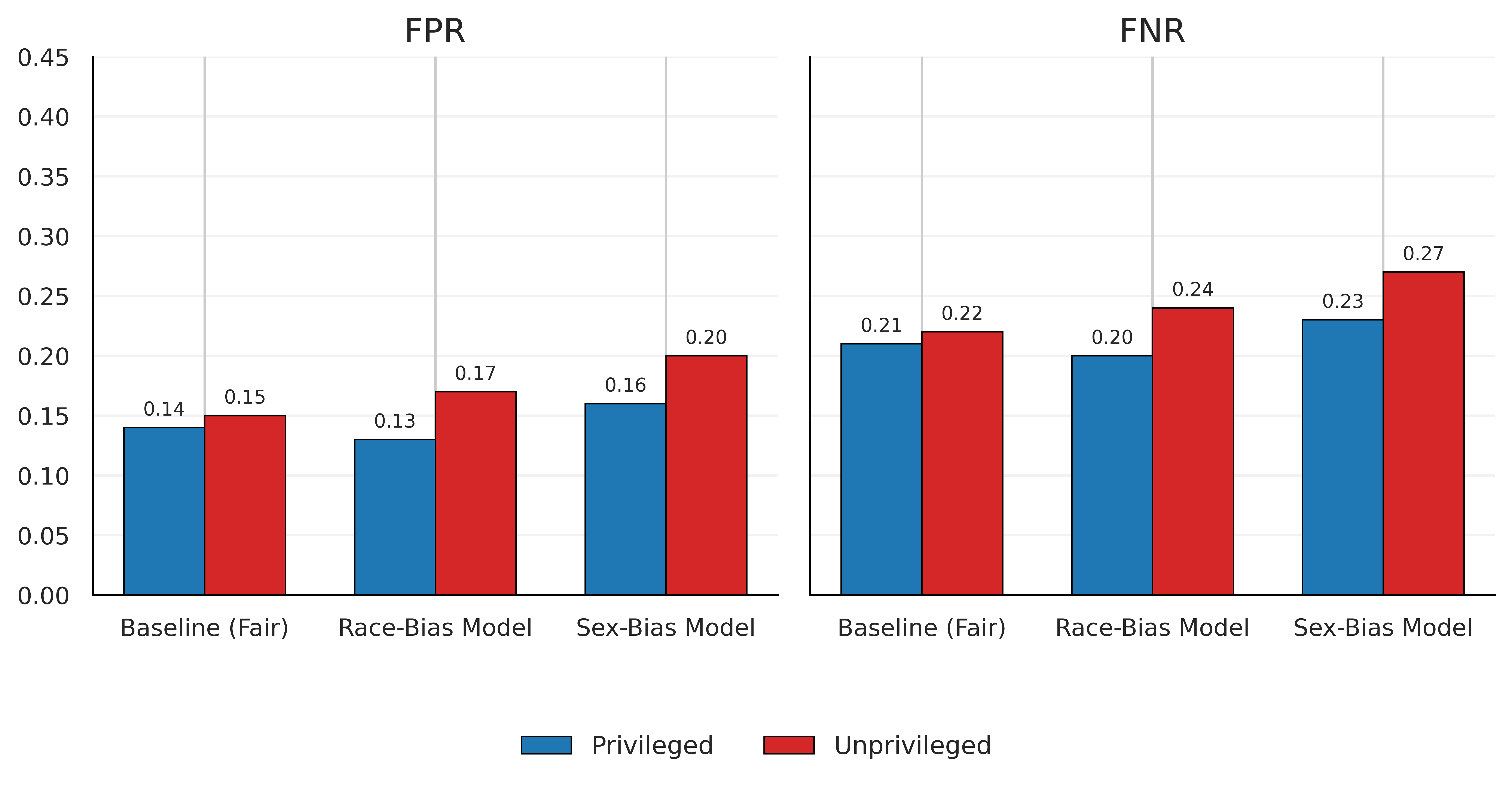}
    \caption{Subgroup misclassification rates by race and gender.}
    \label{fig:sub_race}
\end{figure}

To complement these aggregate fairness metrics, subgroup level performance was analyzed using False Positive Rate (FPR) and False Negative Rate (FNR). Figure~\ref{fig:sub_race} highlights disparities across race and gender subgroups. Pronounced FNR gaps for unprivileged groups indicate systematic under prediction of favorable outcomes, reflecting disparate impact an outcome that the \textit{Nishpaksh} tool appropriately flags as indicative of model bias.

\begin{table}[htbp]
\centering
\caption{Fairness Metric Comparison Across Models}
\label{tab:fairness_metrics}
\begin{tabular}{|c|c|c|c|c|c|}
\hline
\textbf{Metric} & \textbf{Baseline (Fair)} & \textbf{Racial Bias Model} & \textbf{Gender Bias Model} \\ \hline
SPD & 0.187 & 0.106 & -0.287 \\ \hline
NDI & 0.753 & 0.368 & -0.699  \\ \hline
EOD & 0.226 & 0.094 & -0.368   \\ \hline
AOD & 0.176 & 0.074 & -0.273 \\ \hline
EO  & 0.176 & 0.074 & 0.273   \\ \hline
\end{tabular}
\end{table}

\textbf{Summary and key findings:}~The evaluation shows that fairness degradation is very attribute specific and consistent for various fairness metrics. The baseline fair model is closest to ideal case parity and used to validate the effect of fairness constraints. Race and gender bias models exhibit unique fairness deterioration patterns, emphasizing that different sensitive attributes influence fairness in distinct ways. 'Nishpaksh' tool effectively quantifies and interprets these disparities through integrated statistical and model analysis.

\section{Implications for Standardization}
Nishpaksh holds significant implications for standards-driven research and the broader standardization lifecycle, particularly in the Indian context and for emerging technologies like 6G. Firstly, by operationalizing the TEC Standard for AI Evaluation, this work directly addresses a critical post-standardization challenge: bridging the gap between a well-defined national standard and its practical, scalable implementation. The tool provides a concrete mechanism for organizations, including those in the telecom sector, to achieve compliance, conduct audit-grade assessments, and promote responsible AI deployment.

Secondly, Nishpaksh's methodology offers insights for pre-standardization activities. The structured approach to risk quantification via its questionnaire and the systematic evaluation workflow could serve as a template for developing future standards. Specifically, the processes for determining AI application risk grading, attribute-specific bias detection, and standardized reporting have the potential to inform the creation of new technical specifications or industry guidelines, contributing to a more mature IP creation pipeline for global SDOs. This dual contribution underscores Nishpaksh's role not just as a compliance tool, but as a catalyst for advancing the national and international dialogue on trustworthy AI standardization.

\section{Conclusion}
Nishpaksh represents a significant step forward in operationalizing fairness assessment within evolving AI ecosystem. It effectively incorporates analytical and visualization features from global open-source tools while uniquely aligning with the national TEC standard for AI evaluation. The tool's modular architecture and certification-ready reporting capabilities demonstrate its effectiveness in quantifying and interpreting algorithmic bias across multiple dimensions, critical for trustworthy AI deployments in sectors like telecommunications. Experimental validation on the COMPAS dataset confirms that Nishpaksh successfully identifies attribute specific fairness degradation and generates reproducible, auditable assessments, thereby addressing a crucial post-standardization implementation gap. As companies increasingly face pressure from government agencies to ensure algorithmic fairness, Nishpaksh provides an indigenous, implementable solution that enables both self certification and independent third party auditing, positioning it as a foundational tool for responsible AI governance and future standardization discussions. Future versions of Nishpaksh will extend this capability to multimodal AI systems, including models that use text, image and other data types.

\section*{Acknowledgement}
The authors gratefully acknowledge the support of IndiaAI, a Ministry of Electronics and Information Technology (MeitY) initiative, Government of India, for providing funding for this work.

\bibliographystyle{IEEEtran}

\end{document}